\documentclass[prb,twocolumn,showpacs]{revtex4}
\usepackage{amsmath}
\usepackage{amssymb}
\usepackage{bm}
\usepackage{graphicx}

\newlength{\figwidth}
\newlength{\figlarge}
\begin{document}

\title{Quantum condensation in electron-hole bilayers with density imbalance}

\author{Kazuo Yamashita}
\email[Electronic address:]{kazuo@acty.phys.sci.osaka-u.ac.jp}
\affiliation{Department of Physics, Osaka University, Toyonaka, Osaka 560-0043, Japan}  
\author{Kenichi Asano}
\affiliation{Department of Physics, Osaka University, Toyonaka, Osaka 560-0043, Japan}
\author{Takuma Ohashi}
\affiliation{Department of Physics, Osaka University, Toyonaka, Osaka 560-0043, Japan}

\date{\today}

\begin{abstract}
We study the two-dimensional spatially separated electron-hole system with density imbalance at absolute zero temperature. By means of the mean-field theory, we find that the Fulde-Ferrell state is fairly stabilized by the order parameter mixing effect.
\end{abstract}


\maketitle
In the imbalanced systems, where two spices of fermions with unequal densities interacts via interspices attraction, exotic quantum condensations, such as Fulde-Ferrell-Larkin-Ovchinnikov (FFLO) and Sarma phase, are expected to appear.
They are extensively investigated in various research areas such as heavy-fermion, organic superconductors in solid state physics\cite{matsuda}, color superconductivity in high energy physics\cite{casalbuoni} and ultracold fermionic atoms\cite{giorgini}. 
The FFLO phase\cite{fulde, larkin} is characterized by the condensation of the pairs with a finite (nonzero) center-of-mass (COM) momentum.
On the other hand, the Sarma (or breached pair) phase\cite{sarma,liu} can be regarded as mixture of the quantum condensation of composite bosons with zero COM momentum and degenerate fermion gas in the strong coupling regime.
Such a richness of quantum condensation phases in the density imbalanced systems present a great contrast to those in the usual density balanced ones, which show only a crossover between the Bardeen-Cooper-Shrieffer (BCS) condensation in the week coupling regime and the Bose-Einstein condensation (BEC) in the strong coupling regime. 

Since 1960's, a great deal of efforts have been made to observe the quantum condensations of the electron-hole (e-h) pairs in strongly photo-excited semiconductors and semimetallic systems\cite{moskalenko}.
Among them, e-h bilayer systems are considered to be promising, since the radiative lifetimes of the statically separated electrons and holes are long enough to realize thermal equilibrium even at the extremely low temperatures.
In fact, an indication of exciton BEC has already been found experimentally in coupled quantum wells (CQW) for the balanced case\cite{butov}.
They also provide a stage for study on the exotic quantum condensations
in the imbalanced systems, because e-h imbalance can be controlled by
the gates and interlayer bias voltages\cite{imbalance}. 

So far, most of the theoretical works on the quantum condensations of the e-h pairs have been focused on the balanced ones, and the studies on the imbalanced ones still lies far behind. 
Recently, Pieri {\it et al.} have investigated the phase diagram of the e-h bilayer systems with density imbalance at absolute zero temperature\cite{pieri}.
They solve the BCS self-consistent equations only for the Sarma phase, that is only for zero COM momentum of the pair, and identify its instability for the negative superfluid density as the appearance of the FFLO phase.
Although they provide intriguing results, their inequivalent treatment on the FFLO and Sarma phases is insufficient. 

In this paper, we investigate the phase diagrams of the e-h bilayer systems using the mean field approximation, which can take into account the Sarma and FFLO phases on an equal footing.
As will be shown below, the resultant phase diagram is drastically changed: the Sarma phases are strongly suppressed and the FFLO phase occupies relatively high density regions in the phase diagram. 

The Hamiltonian of our reduced model reads
\begin{align}
 H &= \sum_{\bm{k},\sigma} \xi_{\bm{k} \sigma} c^{\dagger}_{\bm{k} \sigma}c_{\bm{k} \sigma} \notag \\
& -\sum_{\bm{k},\bm{k'}} V_{\bm{k},\bm{k'}}c^{\dagger}_{\bm{k}+\bm{q}/2\rm e}c^{\dagger}_{-\bm{k}+\bm{q}/2 \rm h}c_{-\bm{k'}+\bm{q}/2\rm h}c_{\bm{k'}+\bm{q}/2\rm e}, 
\label{eq1}
\end{align}
where $c^{\dagger}_{\bm{k} \sigma}$ creates an electron $(\sigma=\rm e)$ or hole $(\sigma=\rm h)$ with wave vector $\bm{k}$. 
The single particle energy of electrons/holes is expressed as $\xi_{\bm{k} \sigma}=k^2/2m_{\sigma}-\mu_{\sigma}$, and the electron (hole) effective mass and chemical potential are denoted by  $m_{\rm e}$ ($m_{\rm h}$) and $\mu_{\rm e}$ ($\mu_{\rm h}$), respectively. 
We also introduce the interaction matrix element as $V_{\bm k,\bm k'}=S^{-1}\int v(r) e^{i(\bm k-\bm k')\cdot\bm r} d\bm{r}$, where $\bm r$ is the in-plane coordinate, $S$ is the area of the system, and $v(r)$ is the interaction potential for the e-h attraction.
In our Hamiltonian, the electron-electron and hole-hole interactions are ignored, since we expect that their main role is the renormalization of the electron and hole bands.
We also neglect the electrostatic (charging) energy accumulated among the layers and background charge, since such a Hartree contribution gives only a constant energy shift for the fixed electron and hole densities.

Using the mean-field approximation, we obtain the thermodynamic potential at zero temperature as, 
\begin{align}
 \Omega = \sum_{\bm{k}} &\left(\eta^{+}_{\bm{k}}-E_{\bm{k}}\right)+\sum_{\bm{k},\bm{k'}}\Delta_q(\bm{k})\left[V\right]^{-1}_{\bm{k},\bm{k'}}\Delta_q(\bm{k'}) \notag \\
 + &\sum_{\bm{k}}\left[E_{\bm{k}}^+\Theta(-E_{\bm{k}}^+)+E_{\bm{k}}^-\Theta(-E_{\bm{k}}^-)\right],
\label{eq2}
\end{align}
where $E_{\bm{k}}=\sqrt{(\eta^{+}_{\bm{k}})^2+\Delta_q(\bm{k})^2}$, $E_{\bm{k}}^\pm=E_{\bm{k}}\pm\eta^{-}_{\bm{k}}$, $\eta^{\pm}_{\bm{k}}=(\xi_{\bm{k}+\bm{q}/2\rm e}\pm\xi_{-\bm{k}+\bm{q}/2\rm h})/2$, $\Theta(x)$ is the Heaviside step function, and we introduce the order parameter $\Delta_q(\bm{k})$ with the COM and relative momenta $\bm q$ and $\bm k$, respectively.
For simplicity, we assume that the order parameters are real numbers.
Because only the spatially homogeneous order parameters are considered, the FFLO phase discussed here is the ``single-plane-wave'' type, in which the wave function for the COM motion of the e-h pair is a plane wave.
In the following, this type of FFLO phase is called simply FF phase, which is the original one proposed by Fulde and Ferrell\cite{fulde}.

The minimization conditions of the thermodynamic potential give the self-consistent equations, ${\partial\Omega}/{\partial\Delta_q(\bm k)}=0$ and ${\partial\Omega}/{\partial q}=0$, which determine the magnitude of the COM momentum $q$ and the order parameter $\Delta_q(\bm k)$ for the given chemical potentials $\mu_{\rm e}$ and $\mu_{\rm h}$. 
The electron/hole numbers can also be calculated via $N_\sigma = -\partial\Omega/\partial \mu_\sigma$, and the extent of the imbalance is described by the polarization parameter $\alpha =(N_{\rm e}-N_{\rm h})/(N_{\rm e}+N_{\rm h})$.
Solving these self-consistent equations numerically and minimizing thermodynamic potential $\Omega$, we can identify (i) normal phase by $\Delta_q(\bm{k})=0$, (ii) superfluid phase by $\Delta_{q}(\bm{k}) \neq 0$, $q=0$ and $\alpha=0$, (iii) Sarma phase by $\Delta_{q}(\bm{k}) \neq 0$, $q=0$ and $\alpha\neq 0$, and (iv) FF phase by $\Delta_{q}(\bm{k}) \neq 0$, $q\neq 0$ and $\alpha \neq 0$.

We consider both the long- and short-range e-h interactions.
As for the long-range interaction, we use the bare interlayer Coulomb potential, $v(r)={e^2}/{\epsilon\sqrt{r^2+d^2}}$, where $e$ is the electron charge, $\epsilon$ is the background dielectric constant, and $d$ is the interlayer distance.
This choice of interaction is suitable for study of the usual CQW e-h systems, owing to the weak screening effects in the low-dimensional systems.
In this case, we use the two-dimensional (2D) exciton Bohr radius $a_0=\epsilon/2e^2m_{\rm r}$ and binding energy $E_0 = e^2/\epsilon a_0$ as the units of length and energy, respectively. We also set $m_h/m_e=4.3$ and $d=2a_0$ with the typical GaAs/AlGaAs CQW systems in mind.

For the short-range interaction, on the other hand, we use the contact potential, $v(r)= g\delta(\bm r)$.
To avoid the unphysical ultraviolet divergence, the coupling constant $g$ should satisfy
$g^{-1}=S^{-1}\sum_{\bm{k}}(E_a+k^2/2m_{\rm r}-{\rm i}\delta)^{-1}$, where $m_{\rm r}=(m_{\rm e}^{-1}+m_{\rm h}^{-1})^{-1}$ is the reduced mass, $E_a$ is the exciton binding energy, and $\delta$ is a positive infinitesimal.\cite{conduit}
This contact potential is useful to consider the CQW e-h systems where the metallic gate structures are present nearby the quantum wells, since they screen the long-range Coulomb interaction.
Such a consideration is also important because the Hamiltonian (\ref{eq1}) becomes equivalent to that for the two-dimensional (2D) ultracold fermionic atom systems\cite{conduit}. 
Note that the derived self-consistent equations can be much simplified for the contact potential, because the order parameter has no $\bm k$ dependence.
In this calculation, the units of length and energy are chosen as $1/\sqrt{2m_{\rm r}E_a}$ and $E_a$, respectively.

\begin{figure}[t]
\begin{center}
\includegraphics[clip,width=\figwidth]{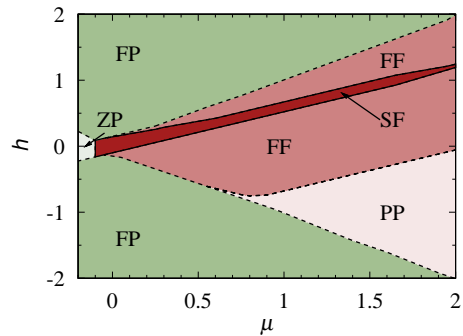}
\end{center}
\caption{
Phase diagram of CQW e-h system in the $(\mu,h)$ plane displaying superfluid(SF), FF, and three normal phases(PP, FP, ZP). Solid, dotted and dot-dashed lines denote the first-order transition, second-order transition and crossover, respectively. 
\label{f1}}
\end{figure}

Figure \ref{f1} shows the $\mu-h$ phase diagram, where $\mu = \left(\mu_e + \mu_h\right)/2 $ and $h = \left(\mu_e-\mu_h\right)/2$. 
For small $\mu$ $(\mu<-0.1)$, there is no electron and hole at $h=0$. 
This phase is referred to as the zero particle (ZP) phase. 
As $|h|$ increases at the fixed value of $\mu$, the fully polarized (FP) normal phase with only electrons($\alpha=+1$), and that with only holes($\alpha=-1$) appear for positive and negative $h$, respectively. 
There is also another normal phase at large $\mu$ $(\mu>0.6)$, which is the partially polarized (PP) normal phase characterized by $|\alpha|\ne 1$.

For $\mu \ge -0.1$, the superfluid (SF) phase appears, which is the quantum condensation without polarization ($\alpha=0$) and has a region with a shape like a ``needle'' in the phase diagram.
This ``needle'' has a positive gradient: the region of the SF phase is shifted to the larger positive side of $h$ with increasing $\mu$. 
This is because the chemical potential of electrons increase more rapidly than that of holes owing to the mass asymmetry $m_{\rm e}<m_{\rm h}$.

One of exotic quantum condensation phases with $\alpha\ne 0$ is the FF phase, which appears in the region around the SF phase.
It is divided by the SF region ($\alpha=0$) into two, that are the electron-rich ($\alpha>0$) and hole-rich ($\alpha<0$) FF phases located at above and below the SF phase, respectively.
The region of the hole-rich FF phase is wider than that of the electron-rich one, because the energy increase due to the reconstruction of the Fermi surfaces by the e-h pair condensation needs more energy for the electrons with light mass than for the holes with heavy mass.
We can also see that the FF phase is much stabilized in the present system than in those with the short-range interaction, e.g. ultracold fermionic atom systems, which exhibit only a narrow FF phase\cite{conduit}.
As discussed bellow, this stabilization can be explained by the order parameter mixing effect.

Another exotic quantum condensation phase with $\alpha\ne 0$ is the Sarma phase. 
This phase is expected to appear in the strong coupling region of the three-dimensional ultracold fermionic atom systems\cite{sheehy,iskin,chien,hu,giorgini,pao,chen}. 
As shown in Fig. \ref{f1}, however, it does not appear in the phase diagram for $m_h/m_e=4.3$, as well as the two-dimensional ultracold fermion systems\cite{conduit,tempere,he}. 
Note that the Sarma phase is unstable for $m_h/m_e<10$ for both long- and short-range interactions. 
However, we find that this phase become thermodynamically stable only in the systems with the extremely large mass ratio, e.g. $m_h/m_e=30$, and with the long-range interaction. 
This is because there is momentum dependence of the order parameter in the case of the long-range interaction, as Forbes {\it et al.} \cite{forbes} claimed before.
\begin{figure}[t]
\begin{center}
\includegraphics[clip,width=\figwidth]{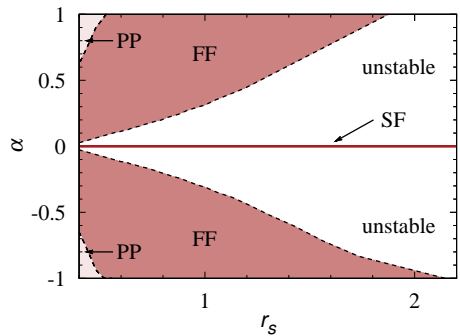}
\end{center}
\caption{
Phase diagram of CQW e-h system in the $(r_s, \alpha)$ plane. The
 abbreviations are the same with those used in Fig.\ref{f1}. 
\label{f2}}
\end{figure}

In order to compare our phase diagram with that obtained by Pieri {\it et al}.\cite{pieri}, 
we redraw the phase diagram on the $(r_s, \alpha)$ plane, where the interaction strength 
is denoted by the dimensionless parameter $r_s=\left[\pi \left(N_e+N_h\right)/2S\right]^{-1/2}$.
In Fig. \ref{f2}, there are the SF phase on the line $\alpha=0$, the FF phase in high density regions, and ``unstable'' regions spread between them.
In these ``unstable'' regions, there is no thermodynamically stable and uniform (single-phase) solution.
We can see that the most of the FF phase and all of the Sarma phase previously found are replaced by these regimes.
This is presumably because only the positiveness of the superfluid
density is checked and the thermodynamic stability of the Sarma phase,
i.e., the positive definiteness of the $2\times2$ matrix $\partial
N_\sigma / \partial \mu_{\sigma'}$,\cite{chen,pao} is not examined in
Ref.~\onlinecite{pieri}.
If we ignore the effects of the background charge, it is expected that the phase separation takes place in the ``unstable'' region, as discussed in the ultracold fermionic atom systems\cite{giorgini,conduit,sheehy,hu,he,bedaque,pao,chen}. 
However, in the realistic CQW e-h system, it hardly occur since the inhomogeneous charge distribution 
give rise to a huge electrostatic energy under the uniform background charge.

\begin{figure}
\begin{center}
\includegraphics[clip,width=\figlarge]{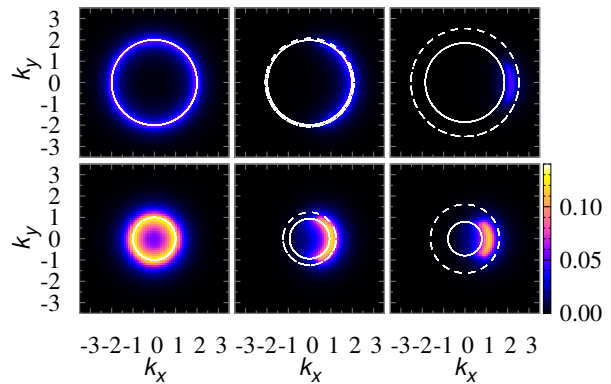}
\caption{
Momentum dependence of the order parameters. 
The upper panels are for $\mu=2.0$ ($r_s \simeq 0.5$) and 
$h=1.24$ ($\alpha=0.0$), $1.19$ ($-0.04$) and $0.79$ ($-0.30$), 
from left to right. 
The lower panels are for $\mu=0.5$ ($r_s \simeq 1.0$) and 
$h=0.31$ ($\alpha=0.0$), $0.21$ ($-0.26$) and $0.01$($-0.61$), 
from left to right. 
Solid and dashed lines are Fermi circle of electrons 
and holes, respectively. 
We choose the center-of-mass momentum as $k_x>0$ and $k_y=0$. 
\label{f3}}
\end{center}
\end{figure}

The FF phase is spread in relatively high density regions and almost all of 
the regions corresponds to the normal phase in the previous study.
The reason for this spreading is that Pieri {\it et al.} studied 
the self-consistent solutions only for the case of zero COM momentum of the pair.
As is clarified below, this change is due to the momentum dependence of 
the order parameter in the case of finite COM momentum of the pair.
In Fig. \ref{f3}, we show the $\bm{k}$-dependence of $\Delta_q(\bm{k})$ for several choices of $r_s$ and $\alpha$. 
In the SF phase with $\alpha=0$, $\Delta_q(\bm{k})$ shows isotropic s-wave behavior forming a circle on the $(k_x, k_y)$ plane. 
As the polarization increases and the system enters the FF phase, the profile of $\Delta_q(\bm{k})$ continuously deforms. 
The region with nonzero $\Delta_q(\bm{k})$ in $\bm{k}$-space shrinks with increasing $\alpha$ and forms a shape of an arc. 
This clearly shows that the order parameter is not the simple s-wave but mixed wave in the FF phase, which is known as ``parity mixing'' or ``order parameter mixing (OPM)'' \cite{matsuo,shimahara}. 
In the simple s-wave FF (sFF) state \cite{fulde,takada}, the total energy of the system is decreased by the condensation but increased by the polarization, because the polarization between sFF phase and normal phase are different for given chemical potentials. 
In contrast, in the mixed FF state, we have confirmed that there is no polarization difference between FF state and normal state, which indicates that there is no energy loss due to the polarization. 

\begin{figure}
\begin{center}
\includegraphics[clip,width=\figwidth]{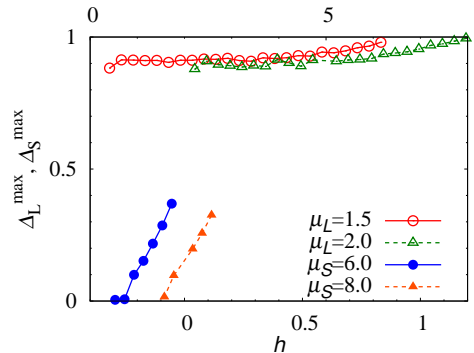}
\end{center}
\caption{$h$ dependence of the maximum values of 
$\Delta_q(\bm{k})$ for the long-range interaction $\Delta_{\mathrm{L}}^{\mathrm{max}}$
 and 
$\Delta_q$ for the short-range interaction 
$\Delta^\mathrm{max}_\mathrm{S}$, 
which are nomalized by 
thir maximum values for each $\mu$.
(The subscript L and S indicate long- and short-range interactions, respectively.)
The upper(lower) scales are for 
the short(long)-range interaction.} 
\label{f4}
\end{figure}

As is shown in Fig. \ref{f4}, we also study the maximum values of $\Delta_q(\bm{k})$ in the FF states for the long- and short- range interactions, which are referred to as $\Delta^\mathrm{max}_\mathrm{L}$ and $\Delta^\mathrm{max}_\mathrm{S}$, respectively.
They are normalized by their corresponding values in the SF phase.
We clearly see that the behavior of $\Delta^\mathrm{max}_\mathrm{L}$ is quite different from those of $\Delta^\mathrm{max}_\mathrm{S}$. 
In fact, $\Delta^\mathrm{max}_\mathrm{S}$ in the SF regime is much larger than that in the FF state. 
As $h$ decreases and the system enters the FF phase ($h<1.8$ for $\mu=6.0$, $h<2.6$ for $\mu=8.0$), $\Delta^\mathrm{max}_\mathrm{S}$ decreases rapidly with increasing the polarization. 
In contrast, in the case of long-range interaction, $\Delta^\mathrm{max}_\mathrm{L}$ is almost unchanged throughout the SF and FF phases due to the OPM effects. 

\begin{figure}
\begin{center}
\includegraphics[clip,width=\figwidth]{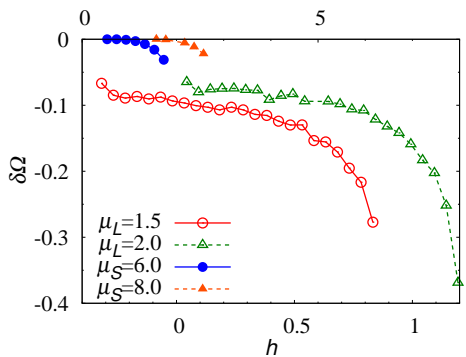}
\end{center}
\caption{
$h$-dependence of $\delta \Omega$ in the FF state.
$\mu_{\mathrm{L}}$($\mu_{\mathrm{S}}$) indicates the mean chemical
 potential of the long(short)-range interaction.
The upper and lower scales are for 
the short- and long-range interactions, respectively. 
\label{f5}}
\end{figure}

Such behavior of the order parameters is directly related to that of the thermodynamic potential. 
We define the thermodynamic potential $\Omega$ measured from the corresponding $\Omega$
 in the normal state as $\delta \Omega = ( \Omega - \Omega_0 ) / \Omega_\mathrm{min}$, 
where  $\Omega_0$ and $\Omega_\mathrm{min}$ are the thermodynamic potential in the 
normal state for given $\mu$ and $h$ and the minimum $\Omega-\Omega_0$ for given $\mu$, respectively. 
In Fig. \ref{f5}, we show the $h$-dependence of $\delta \Omega$ for several choices of $\mu$. 
For the short-range interaction, $\delta \Omega$ almost vanishes in the FF phase, which indicates that the total energy gain in the sFF phase is very small due to the energy loss from the polarization difference and the small energy gain from the small $\Delta^\mathrm{max}_\mathrm{S}$. 
On the other hand, $\delta \Omega$ for the long-range interaction shows significant values in the FF phase, which is about ten times as large as that for the short-range one. 
These results of the order parameters and the thermodynamic potential clearly demonstrate the scenario discussed above. 

We also compare our results with those in the heavy fermion superconductors \cite{matsuo}, where the simple SF and s-p mixed wave FF state are considered.
At zero temperature, the ratio of the critical magnetic field in the s-p mixed FF state to that in the SF state has been obtained as $3.5$. 
In our notations, this means that the ratio of the $h$ range for fixed $r_s$ occupied by the FF phase to that occupied by the SF phase is about $3.5$. 
In our system with the long-range interaction, the ratio is $14$ for $\mu=2.0$ in the electron-rich region ($\alpha>0$).
In this sense, the FF state is much more stable in our system than in the heavy fermion superconductors. 
We can thus conclude that the CQW e-h system is a good candidate for the realization of FF state. 

Finally, we investigate the COM momentum $q$ in the FF phase. 
We find that $q/|k_F^e-k_F^h| \sim 1.2$ at the phase boundary between the SF and FF phase and $q/|k_F^e-k_F^h| \sim 1$ at the boundary between the FF and normal phase. 
The similar behavior also appears in the case of the short-range interaction.
Therefore, this behavior is independent of the range of the interaction and the OPM effects. 
On the other hand, the critical $| \alpha |$ where the phase transition between the FF and normal phase occur is quite different between the cases of the long- and short-range interactions.
The critical $|\alpha|$ is much larger for the long-range interaction than for the short-range one, which is due to the OPM effects.

In conclusion, we investigated the two-dimensional spatially separated electron-hole system with density imbalance at zero temperature.
The results on the previous work\cite{pieri} is almost replaced by the present calculation, particularly in the high density region; The Sarma phase disappers and the FF phase are highly stabilized by the order parameter mixing effect in our system with the long-range interaction. 
We thus consider the CQW e-h system to be a good candidate for the realization of FF state.  

The authors thank T. Ogawa for discussions. 
This work has been supported in part by KAKENHI (No. 20104010, 21740231,
and 21740232) and Global COE Program (Core Research and Engineering of Advanced Materials-Interdisciplinary Education Center for Materials Science), MEXT, Japan.

\end{document}